# Undulation textures at the phase transitions of some alkyloxybenzoic acids.

A. Sparavigna, A. Mello and G. Massa
Dipartimento di Fisica, Politecnico di Torino
C.so Duca degli Abruzzi 24, Torino, Italy

**Abstract**
We observed undulated smectic textures for some compounds of the 4,*n*-alkyloxybenzoic (*n*OBAC) acid series, at transitions between the smectic and the isotropic phase and between the smectic and nematic phase. Studied compounds were 12OBAC, 16OBAC and binary mixtures of 12- and 16OBAC. The undulations are dressing a usual Schlieren texture. An interesting fingerprint pattern is observed at the smectic-nematic in the case of the binary mixture, approximately 1:1 in weight.

## 1. Introduction

Thermotropic liquid crystals are materials possessing one or more mesophases between the isotropic liquid and the solid phase. The mesophase sequence, observed when the temperature is changed, is due to a sequential ordering of molecular arrangements. The shape of molecules is then fundamental to determine the features of this sequence. For rod-like molecules, the material can achieve a nematic phase characterised by an orientational order of molecules. In the smectic phases, the order increases: besides the orientational order, the molecules have an arrangement in layered structure, in some cases with tilt (smectic C) or positional order within the layers (for instance in smectic B) [1-4]. Being strongly anisotropic materials, with optical birefringence properties, a technique commonly applied to detect the material order is the polarized light microscopy, using sandwich cells typically with thickness from 10 to 100 microns. The temperature driven phase transitions are the most beautiful phenomena that can be observed, with sudden changes of colours and textures and very rich in pattern formation.

In this paper we discuss the appearance of undulated textures in the smectic phase of some 4,*n*-alkyloxybenzoic acids, as observed by means of the polarized light microscopy. The undulation instability is observed near the transition temperature at the isotropic-smectic phase transition of 16OBAC and at the nematic-smectic phase transitions of 12OBAC and of some binary mixtures 12OBAC-16OBAC. These textures looks like those observed in the smectic C phase near a smectic C- smectic A transition by Johnson and Saupe [5]. Similar undulations have also been found in the smectic A phase [6,7].

Let us note that the formation of these undulations is a phenomenon different from the formation of the well-known transition bars [4] and must not be confused with. In the case of the binary mixture 12OBAC-16OBAC, transition bars rapidly evolve in fingerprints dressed with stripes.

## 2. The alkyloxybenzoic compounds.

The 4,*n*-alkyloxybenzoic acid (*n*OBAC) compounds have mesophases because their molecules are able to form hydrogen-bonded dimers, rigid and long enough to provide mesogenic conditions. In these compounds, monomers are composed of two sterically distinct molecular parts, the oxybenzoic acid residue and the aliphatic chain. Index *n* is equal to the number of carbon atoms in the aliphatic tail: compounds with *n* ranging from 3 to 6 have a nematic but not a smectic phase. From 7 to 18 carbon atoms in the alkyl tails, the smectic phase appears [8]. The molecular structures of these closed and open dimers have been recently investigated in the framework of ab-initio calculations [9].

Some of these compounds possess, among other interesting features, the texture transition in the nematic phase [10-14]. Optical investigations in compounds with homologous index *n* ranging from 6 to 9, show a nematic phase subdivided in two sub-phases, characterised by different textures. The transition from one nematic sub-phase into the other is ascribed to the growth of cybotactic clusters having short-range smectic order in the nematic phase. Binary mixtures, approximately 1:1 in weight of 6OBAC with other members of the homologous series (7-, 8-, 9-, 12- and 16OBAC), have also the texture transition in the nematic phase. Binary mixtures exhibit an increase in the temperature ranges of smectic and nematic phases, as it happens in other binary mixtures of alkyloxybenzoic acids [15,16]. In the binary mixtures, the mesogenic units are dimers of the same acid (homodimers) but also hydrogen bonded pairs of two different acids (heterodimers) [17]. In spite of the microscopic disorder introduced by mixing two components, the polarised light microscope analysis of the liquid crystal cells reveals the texture transition. This means a persistence of cybotactic clusters, also in the case of mixed dimers.

In this paper we use 12OBAC, 16OBAC and some binary mixtures of them. Let us note that 12OBAC compound does not possess a texture transition in the nematic phase. In fact, the temperature range of nematic phase is quite narrow and at rather high temperature. The smectic-like nematic phase is then suppressed, included in a wide smectic phase. This is due to the fact that the compound has rather long dimers. 16OBAC compound does not possess a nematic phase. The binary mixtures have a narrow nematic phase, accompanied by a wide smectic range.

**3. Sample preparations.**
Prof. M. Petrov of the Institute of Solid State Physics in Sofia, Bulgaria provided us the samples. We used the materials without further purification.

All the compounds were inserted in the cell when the material was in its isotropic phase. The untreated glass surfaces of the cell walls were rubbed with cotton-wool to favour a planar alignment. This procedure for inducing a slight planar alignment was suggested us by researchers of the Liquid Crystal Group at Niopik Institute of Moscow, who referred it as useful in the case of thin cells. The cells we prepared are without spacers and then the thickness of liquid crystal films are of few microns. Previous observations of undulated textures in smectic phases were performed by Johnson and Saupe, using cells where the sample had a thickness ranging form 50 to 150 microns, with the surface of glass slides coated by silane [5], to induce the planar alignment.

In the case of nematic liquid crystals, it is a hybrid configuration favouring the appearance of undulation [18]. Let us remember that a hybrid configuration of a nematic cell means that one of the cell walls is giving a planar alignment and the other a homeotropic alignment. The competition between these two anchoring effects gives rise to threshold phenomena for planar, periodic and hybrid textures. In the case of smectic phase, the origin of the undulations is different. As proposed in Ref.[5], the reason is in the contraction of the layers of the smectic C film. This contraction is expected to be exceptionally large because of the increase of tilt upon cooling. A theoretical model is proposed in Ref.5, in the framework of a Landau approach.

In the case of the thick cells of Johnson and Saupe, the spacers which hold the glasses apart contract less strongly and then mechanical stress results inducing the instability in the smectic texture. In our case, we have no spacers and then the mechanical stress comes from a different origin. It can be due to the presence of transient temperature gradients, arisen in the cell during the cooling: the effects of a different contraction in neighbouring area result in a mechanical stress, which produces the metastable undulation. Moreover the upper glass of our cells is very thin: this glass could play the same role of the spaces in the Johnson and Saupe cells.

We heated and cooled the liquid crystal cells in a Mettler-Toledo thermostage and textures observed with a polarized light microscope. To determine the phase transition temperatures the scanning rate was set to 2 degrees per minute. It is possible to arrest the scanning and held the temperature fixed:

we observed that the instrument displays the temperature fixed within ± 0.1 degrees for approximately one hour.

**4. Undulations at the Isotropic - Smectic transition.**
Some 4,*n*-alkyloxybenzoic acid compounds do not show the nematic phase, because dimers are so long that the smectic C mesophase only is allowed at high temperatures. 16OBAC is one of these compounds; the values of its transition temperatures are given in Table I. If we observe a cell filled with this material with a polarized light microscope, we see the direct transition between isotropic and smectic phases. On cooling, the transition appears as a branched figure growing from the black field of crossed polarizers (see Fig.1). It is also interesting to observe that the material has a smectic texture on cooling, which is different from that displayed on heating. In this case, the texture looks like a nematic texture with schlieren.
In a slow cooling from the isotropic liquid phase, if the temperature is immediately fixed at the instant when the smectic phase starts its formation (132.2°C), undulated textures can be observed near defects (Fig.2). Johnson and Saupe have observed undulation instabilities for the first time, when studying the transition from smectic C to smectic A [5]. As for those observed by Johnson and Saupe, the undulations are "dressing" the texture of the smectic phase and are interpreted as smectic layer undulations, due to the thermal stresses.
In the smectic C of 16OBAC, we see that undulations are metastable, when the sample is kept at the same temperature for enough time. We see that they disappear due to the motion of dislocations: it is the defect line that sweeps out the undulated texture passing through it. This process takes approximately from 5 to 10 minutes, depending on the instability of defects. The sequence of pictures in Fig.3 shows the evolution of undulated texture when the temperature is lowered with a rate of 0.5 degree per minute. From the sequence, it is possible to see that the period of the undulation is increasing. When the temperature is lowered of about two degrees, the dressing texture disappears.

**5. Undulations at the Smectic C – Nematic transition.**
12OBAC compound has a smectic C and a nematic phase. Its transition temperatures are shown in Table I. This material has dimers which are long enough to maintain the smectic order till a high temperature. The nematic phase it is not forbidden, but has a small temperature range. On cooling, the nematic phase appears with coloured nematic bubbles display themselves in the black field of the microscope. Under a further lowering of temperature, the smectic phase appears at 130.2°C. At the nematic-smectic transition, it is possible to observe the transition bars (lower part of Figure 4). The same Figure 4 shows the smectic phase that appears on heating from the crystal phase (on the left) and on cooling from the nematic phase (on the right). In this case, some focal conic domains are possible, with the shape of tears. At lower temperatures, the smectic texture turns out to be more marbled. At 88°C, the material becomes a crystal. At 62°C, we observe a crystal-crystal transition.
On cooling the sample, if the temperature is fixed just 0.5-1 degrees below the nematic-smectic transition, periodic undulations are observable. In Fig.5, we see this kind of texture. It is a texture with a relaxation time of more than half an hour in isothermal conditions. This time is longer than that necessary for the disappearance of undulated textures at the smectic-isotropic transition. This happens because dislocations, which appear at the smectic-nematic transition, are not able to relax and move as freely as they are able to do near the isotropic point. As a consequence there are not any moving dislocations able to remove quickly the undulations. The texture is removed by lowering the temperature of about 5 degrees. In this case again, the behaviour is similar to that of Johnson and Saupe undulations [5].

**6. Binary mixtures.**
We have also investigated several binary mixtures of 12OBAC and 16OBAC. The ratio between weights of 12OBAC and 16OBAC are those shown in Table 1, with the corresponding temperature transitions. The table shows that all these mixtures have a nematic phase. In the case of mixture with ratio 39:61 in weight, sample E, we have the smallest nematic range among the observed ranges. This means that the prevalence of 16OBAC dominates the behaviour of the mixture. Further investigations of the phase diagram are necessary to conclude that the nematic phase is always present.

The mixtures have a clearing point ranging from 134.5 to 138°C. The smectic range, as shown by Table I, is wide, twenty degrees more than that of the single compounds. As previously told, the increase of smectic range in the case of binary mixtures of alkyloxybenzoic acids is an interesting phenomenon, observed for mixtures of 6OBAC with 7-,8-,9-,12- and 16OBAC [13] and other mixtures [15,16]. The binary mixtures investigated in Ref.13 have a texture transition in the nematic phase, transition which is not possessed by the mixtures of 12OBAC and 16OBAC that we have prepared. As in the case of pure 12OBAC, they have a too small nematic range to display such a transition.

Binary mixtures with more 12OBAC can have undulations in the smectic C phase. The texture of the undulation is like that of 12OBAC, shown in Fig.5. Undulations do not appear in sample E, which has a higher content of 16OBAC. Sample A (52:48 in weight) has very beautiful transition bars. These bars have a rapid evolution in a fingerprint patters (Fig.6 on the left). If the temperature is stopped when the fingerprint appears, periodic instabilities dressing the texture can be observed. The behaviour of these instabilities is the same as those observed for 12OBAC. Figure 6 shows the fingerprint texture as it appears at the transition from the nematic to the smectic phase: the fingerprint texture substitutes the transition bars observed in 12OBAC compound. If the temperature decrease is halted as soon as the fingerprints appear, we observe the texture on the right of Fig.6, which clearly shows the dressing of the fingerprint with a periodic pattern.

Transition bars can be observed in the other mixture, but stable fingerprints are shown only by mixture A. As in the case of 12OBAC, the smectic phase can have focal conic domains, which look like tears. For the binary mixture, the tears are surrounded by period patterns and merged in the fingerprints, as is shown in Fig.7. The decrease of temperature destroys the periodic pattern and the fingerprint.

In the binary mixtures, disorder is increased as demonstrated by a wider smectic range. In the nematic phase, the melt is composed of monomers, and open and closed homo- and heterodimers. After transition in the smectic phase, we have a mixture of closed homo- and heterodimers, with different lengths, because of the different number of carbon atoms in monomer tails. This intrinsic disorder increases the smectic range.

The nematic melt of mixtures seems to have not any peculiar property, such as a texture transition or a chiral behaviour. In fact, the mixture 52:48 in weight has a cholesteric-like texture, the fingerprint, that could be due to a chiral-like behaviour of heterodimers, probably the open ones. The chiral-like behaviour is suppressed by a further decrease of temperature, because open dimers turn into closed dimers. Investigation in the framework of ab-initio calculations, as in [9], could be very interesting to find the actual shape of heterodimers.

## 7. Conclusions.

We have prepared some alkyloxybenzoic compounds that show the appearance of an undulated texture, dressing the schlieren texture of their smectic C phase. The behaviour of these undulations is similar that studied by Johnson and Saupe. The binary mixtures have an interesting increase of the temperature range of smectic phase. The intrinsic disorder coming form the presence of homo- and heterodimers is not able to increase the nematic range, which remains rather small.

One of the mixtures, which is approximately 1:1 of components, has a unusual fingerprint texture at the transition between nematic and smectic phases. This cholesteric-like texture appearance could be due to a high percentage of open heterodymers.

**References**
1. P.G.de Gennes and J.Prost: *The Physics of Liquid Crystals* (Clarendon Press, Oxford, 1993)
2. S.Chandrasekhar, *Liquid Crystals* (Cambridge University Press,1992)
3. G.Gray and J.Goodby: *Smectic Liquid Crystals* (Leonard Hill, Glasgow and London, 1984)
4. D. Demus and L. Richter, Texture of Luqid Crystals, VEB Deutscher Verlag fur Grundstoffindustrie, Leizig, 1978.
5. D. Johnson and A. Saupe, Phys. Rev. A **15** 2079 (1977)
6. M. Delaye, G.Ribotta and G.Durand, Phys. Lett. A **44** 139 (1973)
7. N. Clark and R.B. Meyer, Appl. Phys. Lett. **22** 111 (1973)
8. R.F. Bryan, P. Hartley, R.W. Miller and Sheng Ming-Shing, Mol. Cryst. Liq. Cryst. **62** 281 (1980), R. F. Bryan, J. Struct. Chem. **23** 128 (1982).
9. P. Bobadova-Parvanova, V. Parvanov, M. Petrov, L. Tsonev, Crys. Res. Technol. **35** 1321 (2000).
10. M. Petrov, A. Braslau, A.M. Levelut and G. Durand, J. Phys. II (France) **2** 1159 (1992).
11. L. Frunza, S. Frunza, M. Petrov, A. Sparavigna and S.I. Torgova, Mol. Mater. **6** 215 (1996).
12. B. Montrucchio, A. Sparavigna and A. Strigazzi, Liq. Cryst. **24** 841 (1998).
13. A. Sparavigna, A. Mello, B. Montrucchio, Phase Trans. **79** 293 (2006); Phase. Trans. **80**, 191 (2007)
14. A. DeVries, Mol. Cryst. Liq. Cryst. **10** 31 (1970).
15. Ravindra Dhar, R S Pandey and V K Agrawal, Indian J. Pure Appl. Phys. **40** 901 (2002).
16. Seung Koo Kang and E. T. Samulski Liq. Cryst. **27** 371 (2000).
17. A.J. Herbert, Trans. Faraday Soc. **63** 555 (1967).
18. A. Sparavigna, O. D. Lavrentovich and A. Strigazzi, Phys. Rev. E 49 1344 (1994)

TABLE I

| Compound | Transition temperatures (°C) |
|---|---|
| 12OBAC | Cr – 65 – Cr – 91 – Sm C – 131 – N – 138 – I <br> I – 137 – N – 130 – Sm C – 88 – Cr |
| 16OCAC | Cr – 93 – Sm C – 133 – I <br> I – 132 – Sm C – 88 – Cr |
| 12OBAC-16OBAC A (52:48 in weight) | Cr – 71 – Sm C – 132 – N – 138 – I <br> I – 136 – N – 130 – Sm C– 66– Cr |
| 12OBAC-16OBAC B (55:45 in weight) | Cr – 72 – Sm C – 133 – N – 137 – I <br> I – 136 – N – 132 – Sm C – 66 – Cr |
| 12OBAC-16OBAC C (75:25 in weight) | Cr – 69 – Sm C – 130 – N – 136 – I <br> I – 135 – N – 131 – Sm C – 65 – Cr |
| 12OBAC-16OBAC D (82:18 in weight) | Cr – 72 – Sm C – 132 – N – 137 – I <br> I – 136 – N – 128 – Sm C – 66 – Cr |
| 12OBAC-16OBAC E (39:61 in weight) | Cr – 75 – Sm C – 133 – N – 134.5 – I <br> I – 134 – N – 132.5 – Sm C – 66 – Cr |

**Table I:** Transition temperatures (in °C) of alkyloxybenzoic acid compounds on heating and on cooling. 12OBAC and mixtures do not show a texture transition in the nematic phase. 16OBAC exhibits just a smectic phase.

**FIGURES**

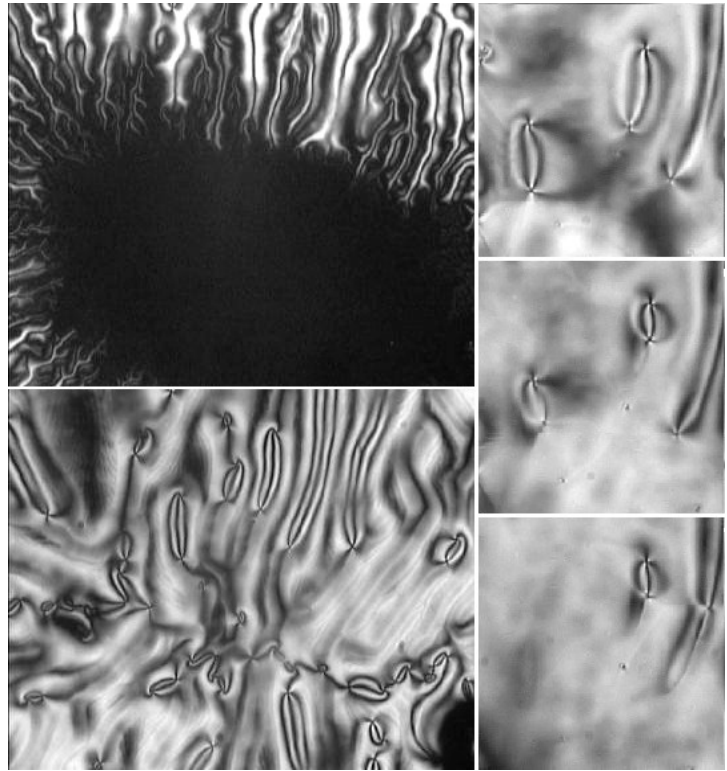

Fig.1 16OBAC compound has a transition directly from the isotropic phase in the smectic C phase. On the right part of the image, the annihilation of defects in the smectic phase. The width of the images on the left is 1 mm.

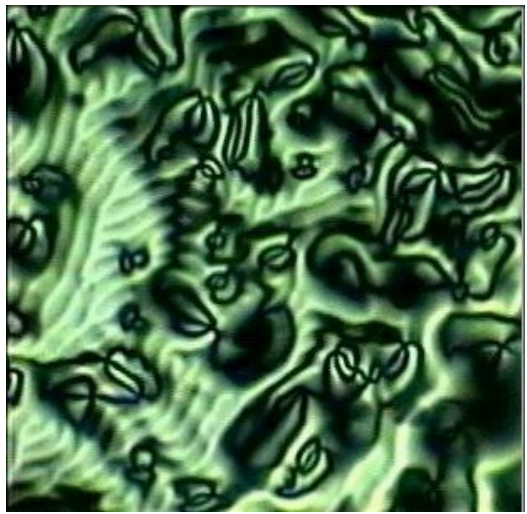

Fig.2 Periodic undulation as it appears near the defects of 16OBAC smectic phase. The undulation is due to stresses on the smectic planes. The image is 0.25 mm width.

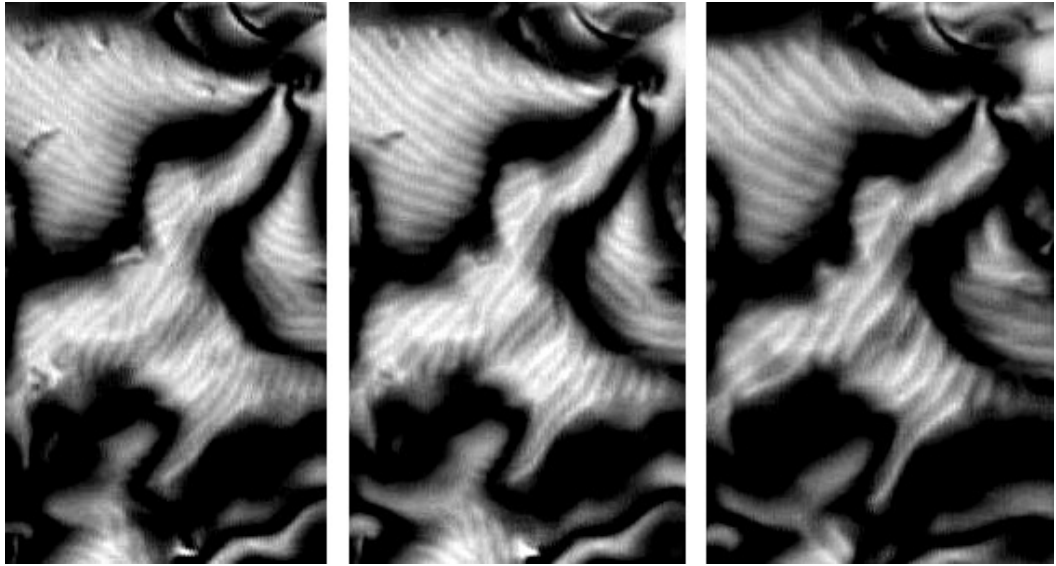

Fig.3 The sequence shows an evolution of the "dressing" texture near defects (16OBAC). From left to right the temperature is lowered of one degree. The stripes are removed by a further decrease of temperature.

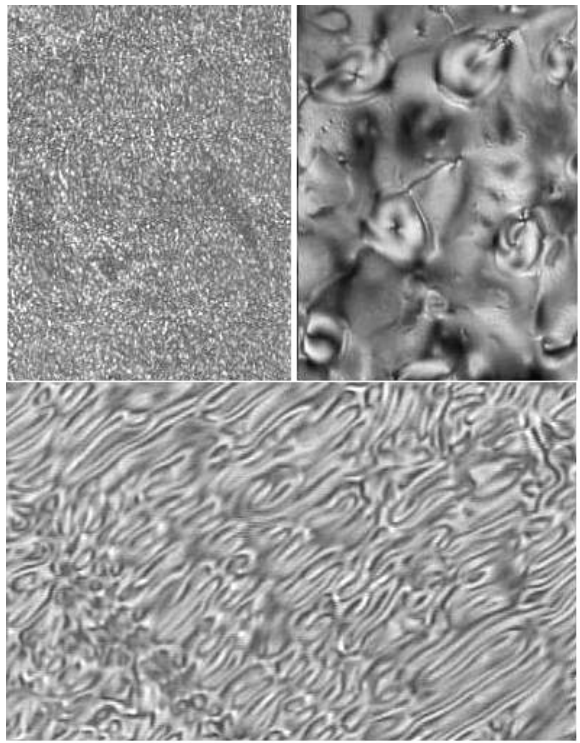

Fig.4 The 12OBAC smectic phase, which appears on heating from a crystal phase (up, on the left) and on cooling from the nematic phase (up, on the right). At the nematic-smectic transition, it is possible to observe the transition bars (down).

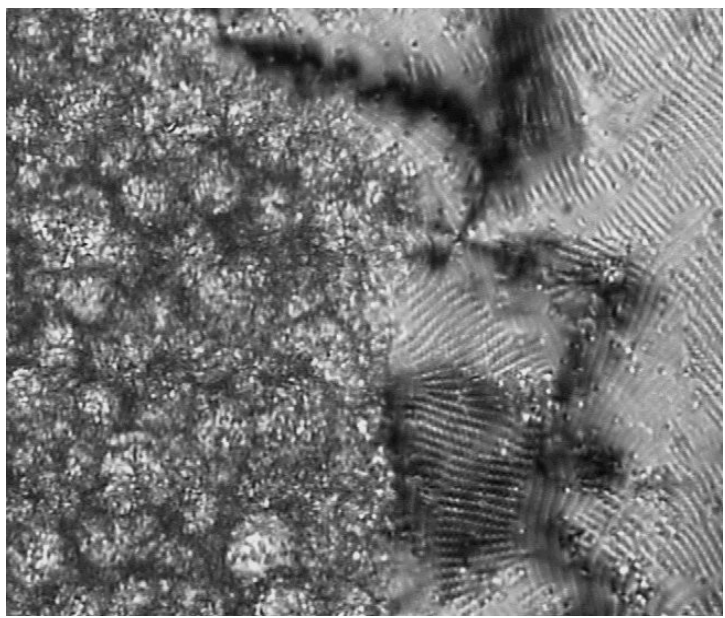

Fig.5 Smectic Schlieren texture with undulation lines, 12OBAC, which appears on cooling from the nematic phase. Image width is 1 mm.

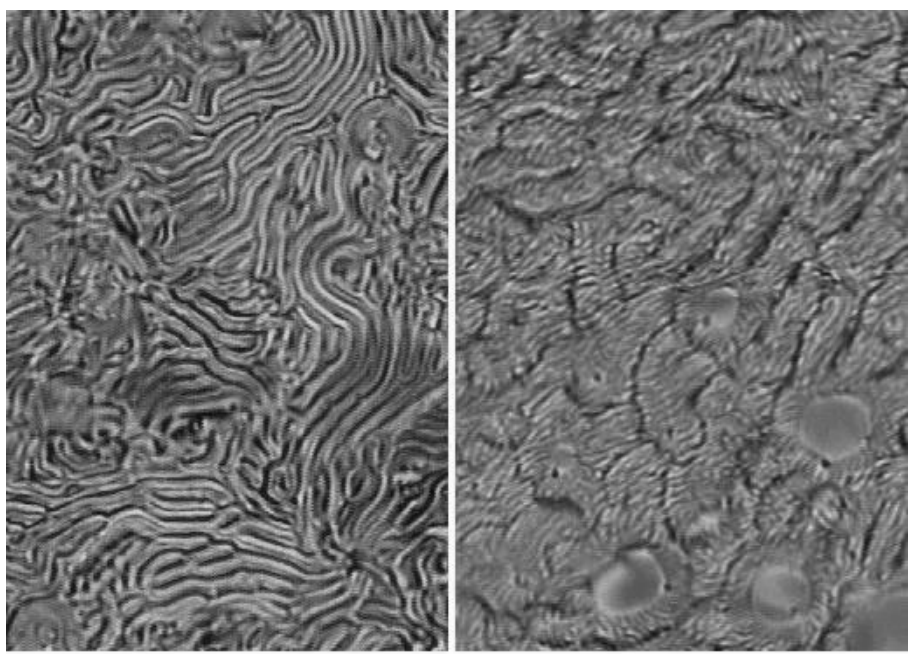

Fig,6 On the left, the fingerprint texture at the transition from the nematic to the smectic phase of the binary mixture 12OBAC-16OBAC. The fingerprint texture substitutes the transition bars observed in 12OBAC compound. Halting the temperature decrease as the fingerprints appear, we observe the texture on the right. Each image has 0.5 mm width.

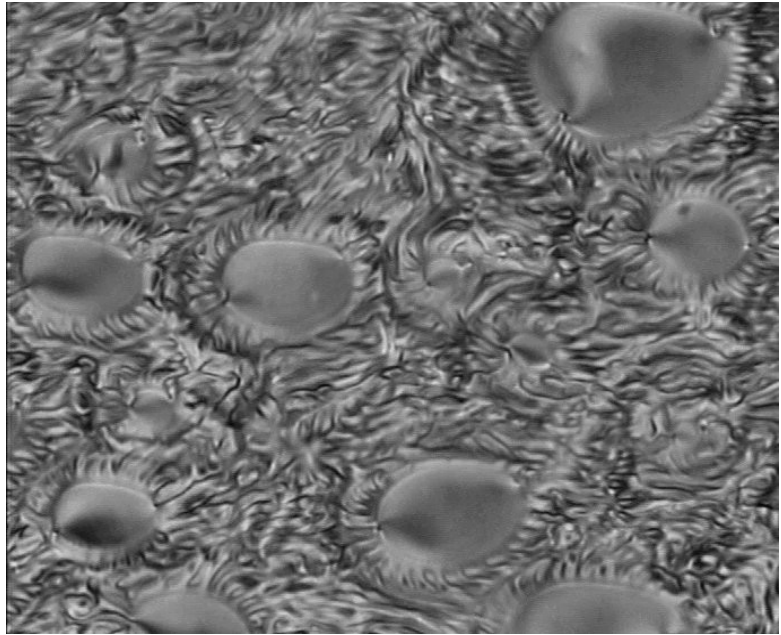

Fig.7 Focal conic domains in the smectic phase of binary mixture 12OBAC-16OBAC. The tears are surrounded by period patterns and merged in the fingerprints. The image width is 1 mm.